\documentclass[a4paper]{article}
\usepackage{interspeech2009,amssymb,amsmath,epsfig}
\usepackage{graphicx}
\ninept
\def\reg{{\rm\ooalign{\hfil
     \raise.07ex\hbox{\scriptsize R}\hfil\crcr\mathhexbox20D}}}

\title{A Deterministic plus Stochastic Model of the Residual Signal for Improved Parametric Speech Synthesis}

\makeatletter
\def\name#1{\gdef\@name{#1\\}}
\makeatother
\name{{\em Thomas Drugman $^1$, Geoffrey Wilfart $^2$, Thierry Dutoit $^1$}}

\address{$^1$ TCTS Lab, Facult\'e Polytechnique de Mons, Belgium \\
$^2$ Research and Development, Acapela Group, Mons, Belgium\\
{\small \tt thomas.drugman@fpms.ac.be}}

\begin{document}
\maketitle

\begin{abstract}
Speech generated by parametric synthesizers generally suffers from a typical buzziness, similar to what was encountered in old LPC-like vocoders. In order to alleviate this problem, a more suited modeling of the excitation should be adopted. For this, we hereby propose an adaptation of the Deterministic plus Stochastic Model (DSM) for the residual. In this model, the excitation is divided into two distinct spectral bands delimited by the maximum voiced frequency. The deterministic part concerns the low-frequency contents and consists of a decomposition of pitch-synchronous residual frames on an orthonormal basis obtained by Principal Component Analysis. The stochastic component is a high-pass filtered noise whose time structure is modulated by an energy-envelope, similarly to what is done in the Harmonic plus Noise Model (HNM). The proposed residual model is integrated within a HMM-based speech synthesizer and is compared to the traditional excitation through a subjective test. Results show a significative improvement for both male and female voices. In addition the proposed model requires few computational load and memory, which is essential for its integration in commercial applications.
\end{abstract}



\section{Introduction}\label{sec:intro}
Statistical parametric speech synthesizers have recently shown their ability to produce natural-sounding voices \cite{Black},\cite{Blizzard}. Compared to other state-of-the-art techniques, they present the advantage of being flexible while requiring a low footprint. This makes them particularly suited for small devices. On the other side, their main drawback lies in the buzziness of the delivered quality, as typically encountered in LPC-like vocoders. This can be mostly explained by the parametrical representation the synthesizer relies on. While methods capturing the spectral envelope have nowadays reached maturity, there still is a lot to be gained in finding a suited modeling of the excitation.

The traditional excitation used by HMM-based speech synthesizers is either a pulse train or a white noise, during voiced and unvoiced segments respectively. In order to enhance this model, some approaches have been proposed in the literature. In \cite{Yoshimura}, Yoshimura et al. integrated the Mixed Excitation (ME) coding method. In this framework, the excitation is obtained using a multi-band mixing model, containing both periodic and aperiodic contributions, and controled by bandpass voicing strengths. In a similar way, Maia et al. \cite{Maia} made use of high-order filters to obtain these components. In \cite{Cabral07} and \cite{Cabral08}, Cabral et al. suggested the integration of a Liljencrants-Fant waveform (\cite{LF}) as a modeling of the glottal source. In \cite{Drugman} we proposed the use of a codebook of pitch-synchronous residual frames to construct the voiced excitation. All these techniques tend to relatively reduce the produced buzziness, and therefore improve the overall quality.

The main motivation of this paper arises from the unsatisfying results we obtained on female voices in \cite{Drugman}. Indeed while this latter method led to a convincing improvement over the traditional excitation for the male speakers, subjective tests were more mitigated for the female voices. For this, we propose an adaptation of the Deterministic plus Stochastic Model (DSM) for the residual signal. The most famous example of DSM is the Harmonic plus Noise Model (HNM) which already showed its high-quality parametrization of speech \cite{Stylianou},\cite{YannisPhD}. According to this model, speech is composed of a low-frequency harmonic structure and a high-frequency noise, assumed to principally model the turbulences of the glottal airflow. The spectrum is then divided into two bands delimited by the so-called \emph{maximum voiced frequency} $F_m$. Our idea in this paper is to adopt a similar approach on the residual signal. While the stochastic part may stay unchanged, some modifications have to be brought to the deterministic component. For this, similarly to \cite{Drugman}, we propose to model the low-frequency contents (below $F_m$) by decomposing pitch-synchronous residual frames on an orthonormal basis obtained by Principal Component Analysis (PCA). The resulting DSM then consists of a compact representation of the residual, requiring few computational load and memory, which make it suited for its integration within small devices.

The paper is structured as follows. Section \ref{sec:DSM} details the way our pitch-synchronous residual frames are obtained and how they are modeled by the Deterministic plus Stochastic Model (DSM). Section \ref{sec:HTS} presents the incoporation of the DSM in our HMM-based speech synthesizer. The traditional and proposed excitations are then compared by a subjective test in Section \ref{sec:experiments}. Finally Section \ref{sec:conclu} concludes.

\section{A Deterministic plus Stochastic Model of pitch-synchronous residual frames}\label{sec:DSM}

This Section first describes how the pitch-synchronous residual frames are obtained from speech signals (\ref{ssec:residual}). These frames have the particularity to be centered on a Glottal Closure Instant (GCI), two period-long and Blackman-windowed. This processing makes them comparable so that they are suited for a common modeling. For this, we apply a DSM (\ref{ssec:DSM}) consisting of (a) a low-frequency deterministic model, based on a PCA decomposition (\ref{sssec:deterministic}), and (b) a high-frequency modulated noise component (\ref{sssec:stochastic}).
 
\subsection{Pitch-synchronous residual frames}\label{ssec:residual}

\begin{figure}[!ht]
  \centering
  \includegraphics[width=0.45\textwidth]{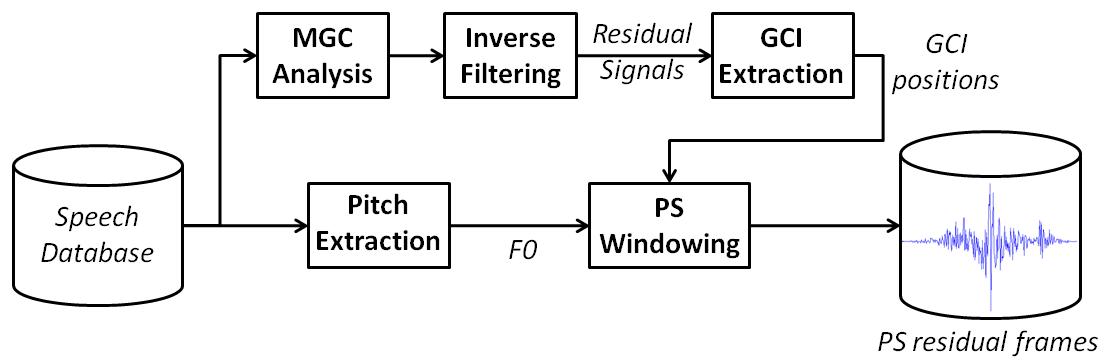}
  \caption{Workflow for obtaining the pitch-synchronous residual frames.}
  \label{fig:PSframes}
\end{figure}

The workflow for obtaining the pitch-synchronous residual frames is presented in Figure \ref{fig:PSframes}. First a Mel-Generalized Cepstral (MGC) analysis is performed on the speech signals, as these features have shown their efficiency to capture the spectral envelope \cite{MGC}. As recommended in \cite{Blizzard}, we used the parameter values $\alpha=0.42$ ($Fs=16kHz$) and $\gamma=-1/3$ for the MGC extraction. Residuals are then obtained by inverse filtering. Glottal Closure Instants (GCIs) are then identified by locating the greatest discontinuity in the residual signal as explained in \cite{DrugGCI}. In parallel, the pitch is estimated using the publicly available Snack Sound Toolkit \cite{Snack}. The pitch-synchronous residuals are finally isolated by a GCI-centered and two period-long Blackman windowing.

\subsection{The proposed Deterministic plus Stochastic Model}\label{ssec:DSM}

As previously mentioned, the DSM consists of the superposition of a deterministic $r_d(t)$ and stochastic $r_s(t)$ components of the residual. These components act in two distinct spectral bands delimited by the frequency $F_m$, equivalent to the \emph{maximum voiced frequency} in the HNM model. The DSM we applied on the pitch-synchronous residual frames is described here below.

\subsubsection{The maximum voiced frequency}\label{sssec:Fm}

The maximum voiced frequency demarcates the boundary between both deterministic and stochastic components. Although some methods have been proposed for its estimation \cite{Stylianou}, we use in this work a fixed value at 4 kHz, as done in \cite{YannisPhD} or \cite{Pantazis}.

\subsubsection{The deterministic modeling}\label{sssec:deterministic}
\begin{figure}[!ht]
  \centering
  \includegraphics[width=0.4\textwidth]{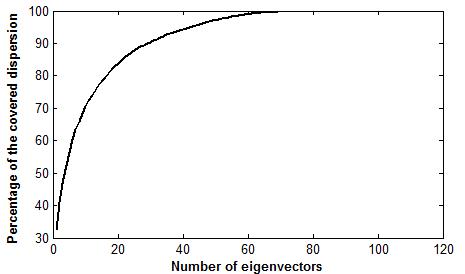}
  \caption{Evolution of the relative dispersion covered with an increasing number of eigenvectors.}
  \label{fig:EigenVal}
\end{figure}

\begin{figure}[!ht]
  \centering
  \includegraphics[width=0.4\textwidth]{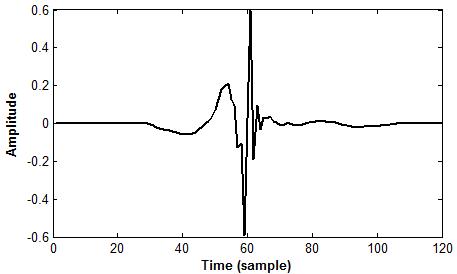}
  \caption{The first eigenvector for speaker SLT.}
  \label{fig:EigenVec}
\end{figure}

In order to model their low-frequency contents, we propose to decompose pitch-synchronous residual frames (as extracted in \ref{ssec:residual}) on an orthonormal basis obtained by PCA. For this, a dataset of such frames is constructed, and frames are normalized in both pitch and energy. This step ensures the coherence of the dataset before applying PCA. Assuming the residual signal as an approximation of the glottal source, resampling the residual frames by interpolation and decimation should preserve their shape and consequently their most important features. Similarly to \cite{Drugman}, care has to be taken when choosing the number of points for the length-normalization. In order to avoid the appearance of energy holes at synthesis time (occuring if the useful band of the deterministic part does not reach $F_m$), the pitch value $F_0^*$ for the normalization is such that:

\begin{equation}\label{eq:pitch}
F_0^* \leq \frac{F_N}{F_m} \cdot F_{0,min}
\end{equation}

where $F_N$ and $F_{0,min}$ respectively denote the Nyquist frequency and minimum pitch value for the considered speaker. PCA can now be calculated on the dataset allowing dimensionality reduction and feature decorrelation. The number of retained eigenvectors is chosen such that they explain around $80\%$ of the total dispersion. Through unformal Analysis/Synthesis experiments, this value was reported to give almost inaudible degradations. Besides we observed that once the dataset for PCA computation reached about 10 minutes, extracted eigenvectors remained sensibly unchanged. Figure \ref{fig:EigenVal} shows the evolution of the covered relative dispersion with the number of eigenvectors. In this case, using 15 features is sufficient for giving high-quality coding results. For information, the first eigenvector is exhibited in Figure \ref{fig:EigenVec}. A strong similarity with the LF model \cite{LF} can be noticed.

\subsubsection{The stochastic modeling}\label{sssec:stochastic}

The stochastic modeling of the residual $r_s(t)$ that we adopted is identical to the noise part in the HNM model \cite{Stylianou}. It corresponds to a white Gaussian noise $n(t)$ convolved with an auto-regressive model $h(\tau,t)$ and whose time structure is controled by a parametric envelope $e(t)$:

\begin{equation}\label{eq:noise}
r_s(t)=e(t)\cdot[h(\tau,t)\star n(t)]
\end{equation}

\begin{figure*}[!ht]
  \centering
  \includegraphics[width=0.95\textwidth]{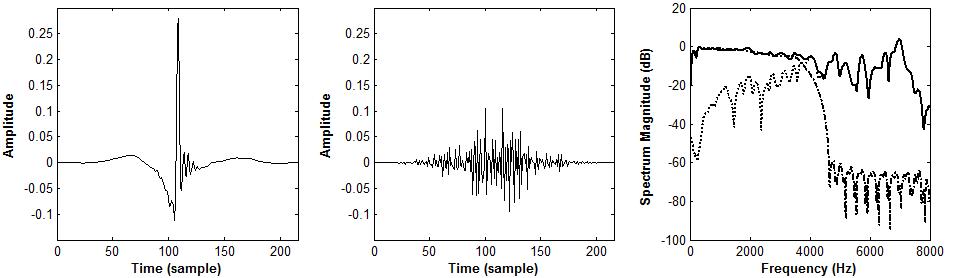}
  \caption{Example of DSM decomposition on a pitch-synchronous residual frame. \emph{Left panel:} the deterministic part. \emph{Middel panel:} the stochastic part. \emph{Right panel:} amplitude spectra of the deterministic part (dashdotted line), the noise part (dotted line) and the reconstructed excitation frame (solid line) composed of the superposition of both components.}
  \label{fig:Decomp}
\end{figure*}

Although some studies focus on a better modeling of the envelope $e(t)$ \cite{Pantazis}, we use in this work a pitch-dependent triangular function as suggested in \cite{YannisPhD}. Furthermore, since $F_m$ is fixed to 4 kHz in our case and that the residual spectrum is almost flat over the whole frequency range, it is reasonable to consider that the auto-regressive model has the same effect on all the frames: it acts as a high-pass filter (beyond $F_m$) slightly attenuated in the very high frequencies. Consequently $h(\tau,t)$ is computed once and for all for the rest of this work. An example of DSM decomposition is displayed in Figure \ref{fig:Decomp}.

\subsubsection{The DSM vocoder}\label{sssec:vocoder}

\begin{figure}[!ht]
  \centering
  \includegraphics[width=0.45\textwidth]{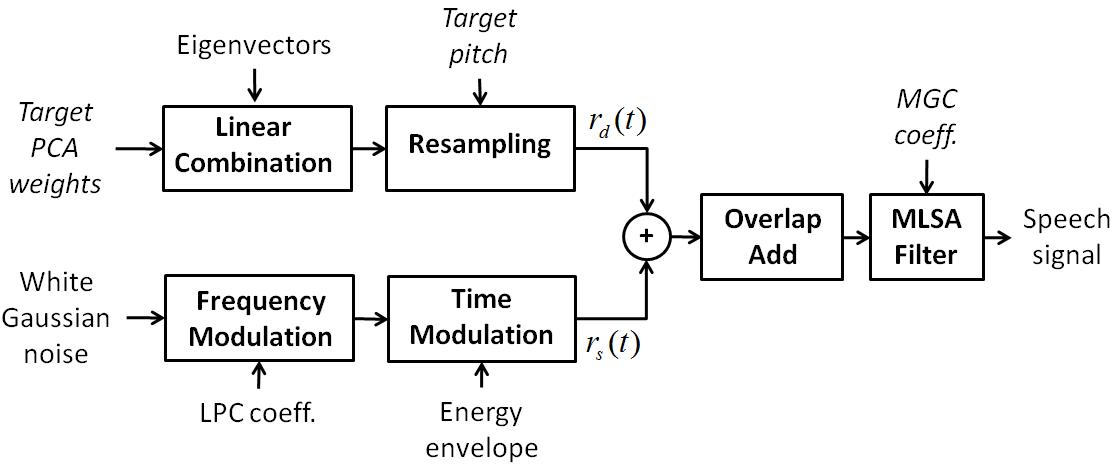}
  \caption{Workflow of the DSM vocoder.}
  \label{fig:Vocoder}
\end{figure}

A workflow summarizing the DSM vocoder can be found in Figure \ref{fig:Vocoder}. Inputs are the PCA weights, the pitch and the MGC coefficients. A first version of the deterministic component of the residual $r_d(t)$ is obtained by a linear combination of the eigenvectors. The resulting waveform is resampled such that its length is twice the target pitch period. Following Equation \ref{eq:noise}, the stochastic part $r_s(t)$ is a white noise modulated by an AR model and multiplied by a triangular envelope centered on the current GCI. Both components are superposed and then overlap-added so as to obtain the residual signal. Note that in the case of unvoiced regions, the excitation consists of a simple white Gaussian noise. The residual is finally the input of the Mel-Log Spectrum Approximation (MLSA) filter to get the speech signal.

\section{Integration in the HMM-based speech synthesizer}\label{sec:HTS}

The implementation of our HMM-based speech synthesizer relies on the HTS toolkit publicly available in \cite{HTS}. Some modifications are nonetheless required to incoporate the proposed excitation. For this, three streams of data are considered: one stream for the MGC coefficients, one for the pitch, and one for the PCA weights of the deterministic part. For the last two streams, one needs to distinguish between voiced and unvoiced regions, which led us to adopt a multi-space distribution (MSD), as described in \cite{MSD}. Following \cite{GenHMM}, the first derivatives of each stream are included into our model, so that the configuration is the following:

\begin{itemize}
\item one single Gaussian distribution with diagonal covariance for MGC coefficients and their derivatives.
\item one MSD distribution for pitch, 
\item one MSD distribution for pitch derivatives,
\item one MSD distribution for PCA weights,
\item one MSD distribution for PCA weight derivatives.
\end{itemize}

In each MSD distribution, for voiced parts, parameters are modeled by single Gaussian distributions with diagonal covariance, while the voiced/unvoiced decision is modeled by an MSD weight. At synthesis time, parameters generated from a constrained maximum likelihood algorithm, as described in \cite{GenHMM}, are fed into our DSM vocoder to produce the synthetic speech. We have observed that, in the generated residuals, the contribution of the first eigenvector clearly dominates the deterministic component. Listening tests confirmed that using eigenvectors of superior ranks led to no audible difference. For this reason, experiments presented in Section \ref{sec:experiments} were carried out using only the first eigenvector. Under this assumption, excitation is only characterised by the pitch, and the MSD stream of PCA weights may be removed. This leads to a very simple model, in which the voiced excitation essentially (below $Fm$) consists of a single waveform that is resampled to the target pitch period, requiring almost no computational load, while providing high-quality synthesis.

\section{Synthesis experiments}\label{sec:experiments}

The synthetic voices of five speakers were assessed: AWB (Scottish male), Bruno (French male), Julie (French female), Lucy (US female) and SLT (US female). AWB and SLT come from the publicly available CMU ARCTIC database \cite{ARCTIC} and about 45 minutes of speech for each were used for the training. Other voices were kindly provided by Acapela Group and were trained on a corpus of around 2 hours. The test consists of a subjective comparison between the proposed and the traditional pulse excitation. For this, 40 people participated to a Comparative Mean Opinion Score (CMOS,\cite{CMOS}) test composed of 20 randomly chosen sentences of about 7 seconds. For each sentence they were asked to listen to both versions (randomly shuffled) and to attribute a score according to their overall preference (see Table \ref{tab:CMOS}).

\begin{table}[!ht]
\centering
\begin{tabular}{c | c}
\hline
\hline
Much better & +3 \\
\hline
Better & +2 \\
\hline
Slightly better & +1 \\
\hline
About the same & 0 \\
\hline
Slightly worse & -1 \\
\hline
Worse & -2 \\
\hline
Much worse & -3 \\
\hline
\hline
\end{tabular}
\caption{Grades in the CMOS scale. The reference signal used either the traditional or proposed excitation.}
\label{tab:CMOS}
\end{table}

\begin{figure}[!ht]
  \centering
  \includegraphics[width=0.45\textwidth]{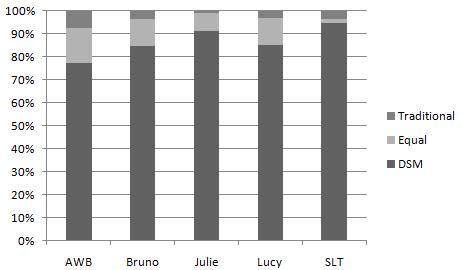}
  \caption{Preference score for the five speakers.}
  \label{fig:Pref}
\end{figure}

\begin{figure}[!ht]
  \centering
  \includegraphics[width=0.44\textwidth]{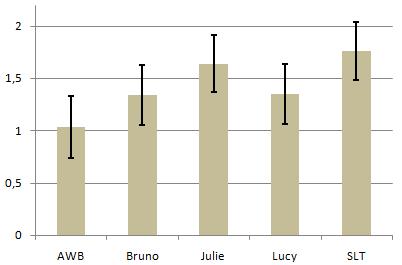}
  \caption{Average CMOS score in advantage of the DSM for the five speakers, together with their 95\% confidence interval.}
  \label{fig:CMOS}
\end{figure}

Preference scores can be viewed in Figure \ref{fig:Pref}. A clear improvement over the traditional pulse excitation can be observed for all voices. Compared to the method using a pitch-synchronous residual codebook proposed in \cite{Drugman}, results are almost similar on male speakers, with a minor loss of less than 5\% for both AWB and Bruno. On the contrary, the contribution on female voices is much more evident. While only 30\% of participants prefered the technique using the codebook for speaker SLT \cite{Drugman}, score now reaches more than 90\% for the DSM. This trends holds for other female voices. Figure \ref{fig:CMOS} exhibits the average CMOS scores with their 95\% confidence intervals. Average values vary between 1 and 1.75, confirming a clear advantage for the proposed technique. 

\section{Conclusion}\label{sec:conclu}
This paper proposed an adaptation of the Deterministic plus Stochastic Model for residual signals. This approach is motivated by the need of finding an efficient and compact representation of the excitation, so as to reduce the buzziness produced by parametric speech synthesizers. Through subjective tests, the method was shown to clearly outperform the state-of-the-art excitation for five speakers. In addition the mitigated results we obtained for female voices in a previous work are now undeniably overcome. In the same time, the method preserves the small footprint as well as the weak computational requirements, making it suited for its integration within small devices.

\section{Acknowledgments}\label{sec:Acknowledgments}

Thomas Drugman is supported by the ``Fonds National de la Recherche
Scientifique'' (FNRS). Authors also would like to thank the participants of the subjective test.

\eightpt
\bibliographystyle{IEEEtran}

\end{document}